\documentstyle[sao2,psfig]{article}
\issue{1998, 44, 110--118}
\begin{document}
\title{Modelling emission line profiles of a non--uniform accretion disk}
\author{N.V. Borisov \and V.V. Neustroev}
\institute{\saoname}
\date{September 26, 1997}{November 13, 1997}
\maketitle
\begin{abstract}

Techniques of calculation of emission line profiles formed in a non--uniform
accretion disk is presented. Change of the profile shape as a
function of  phase turn of the disk with a bright spot on the surface is
analysed. A possibility of a unambiguous determination of the disk
and spot parameters is considered, the accuracy of their determination is
estimated.  The results
of calculations show that the analysis of spectra obtained at different
phases of the orbital period gives a possibility of estimating the basic
parameters of the spot (such as geometric size and luminosity) and to
investigate the structure of the accretion disk.  

\end{abstract} 

\section{Introduction} 
By the present time a large number of
close binary systems containing a component with an accretion disk have been
detected. In such systems a secondary nondegenerate star fills its critical
Roche lobe and transfers matter to the primary star through the inner
Lagrangian point mostly as a gas stream. Due to the high angular momentum the
outflowing gas forms the accretion disk around the primary ``peculiar''
component. At the place  the gas stream strikes the outer rim of the disk an
area of enhanced temperature and luminosity, named bright spot, is formed.

In Algol--type systems the disk accretion occurs on a normal B--A
main--sequence star (Plavec, 1980), and in cataclysmic variables and X--ray
binaries it occurs on a degenerate star (Kraft, 1965). In the last
case the
accretion disk may give an appreciable contribution to the optical
continuum (Pringle and Rees, 1972; Shakura and Sunyaev, 1973).

Cataclysmic variables are the most suitable objects for the study of accretion
disks. These close binary systems consist of a
white dwarf (primary) and a main--sequence star of a late spectral class
(G--M). The
choice of cataclysmic variables for research into accretion disks is
defined by the following factors:
\begin{enumerate}\item The main part of energy radiation from the disk is
emitted at optical and
ultraviolet wavelengths;
\item The contribution of the secondary component (a star of the late
spectral type G--K) to the system total luminosity is relatively small in
comparison with that of the accretion disk;
\item Cataclysmic variables are usually  more amenable          to
observations than low--mass X--ray binaries. This makes it definitely easier
to get observational data of high quality; \item The number of cataclysmic
variables is much greater than the number of representatives of other types
of binaries with accretion disks.  \end{enumerate}

Typical optical spectra of cataclysmic variables contain emission
lines of hydrogen, neutral helium and singly ionized calcium all superposed
onto a blue continuum. He\,II 4686 may also be present. In the spectra of
systems with high inclinations the emission lines of H and He\,I are
usually  double--peaked and have profiles with base widths over
2000--3000 km/s (Williams, 1983; Honeycutt et al., 1987).

The double--peaked profile is a result of Doppler shift of emission from
the accretion disk (Smak, 1969; Horne and Marsh, 1986). The profiles are
often observed to be asymmetric and the intensities of the red and blue
peaks are variable
with the orbital period phase (Greenstein and Kraft, 1959). The trailed
spectra show
strong double--peaked symmetrical line profiles and a weak narrow
component
which forms a S--wave  due to sinusoidal variations
of its radial velocity. The ``S-wave'' component is usually attributed to
the bright spot --- the point of interaction of the gas stream and the
accretion disk (Kraft, 1961; Smak, 1976), moreover their physical parameters
define the nature of the processes involved (Livio, 1992). This makes it
important to study the observational properties in the investigation of
accretion in close binary systems.

In this paper we consider a method of modelling emission line profiles
which are formed in non-uniform accretion disks. We study the
line profile variation
depending on the phase turn of the disk with the bright spot on
the surface. In section 2 the model and the technique of calculations are
described.  In
section 3 we test this method for the determination of the line profile
from the model parameters. Evaluation of the accuracy of
determination of parameters is described in section 4. Finally, in section 5
the results of calculations are given.

\section{Method}

A powerful tool in the
investigation of the orbital variations of emission lines in spectra of close
binaries, which is widely used now, is Doppler tomography. This is an
indirect imaging technique which can be used to determine the
two--dimensional velocity-field distribution of line emission in binary
systems (Marsh and Horne, 1988). This method provides very accurate
reconstructions even when analyzing low S/N--ratio spectra. However, such a
powerful tool in  studying the structure of  accretion disks is
unfortunately not free from demerits. We point out only the basic of them.
\begin{itemize} \item The computation of a Doppler map requires a large
number of high resolution spectra covering an orbital period. This is a
problem when studying weak and short--period close binary systems.
\item
The variation in flux of the disk details during observations may distort
the map (Marsh and Horne, 1988).  This occured, for example, in the
study of the cataclysmic variable U Gem (Marsh et al., 1990).  In the Doppler
map the ring--shaped component from the disk is weakened near the bright
spot. This weakening is actually non--existent.  \item
Since all the observational data are used for computation of the map, we
lose the possibility of studying variations of fluxes from the bright spot
and other emission regions over the orbital period.  \end{itemize}

The enumerated shortcomings of the method of Doppler tomography restrict
possibilities of
its use. The modelling of the line profiles obtained at different phases of
the
orbital period is another possible method of analysis of variations of the
accretion disk and  spot parameters with time, without
mapping the disk in the velocity field. Thus, at a sacrifice the high
spatial
resolution we obtain a possibility of studying  the temporal variability.

For accurate calculation of line profiles formed in the accretion disk, it is
necessary to know the velocity field of radiating gas, its temperature and
density, and, first of all, to calculate the radiative transfer equations in
lines and the balance equations. Unfortunately, this
complicated
problem has not been solved until now and it is still not possible to reach an
acceptable consistency between calculations and observations. Nevertheless,
even the simplified models allow one to define some important
parameters of the accretion disk.  \begin{figure}
{\vbox{\psfig{figure=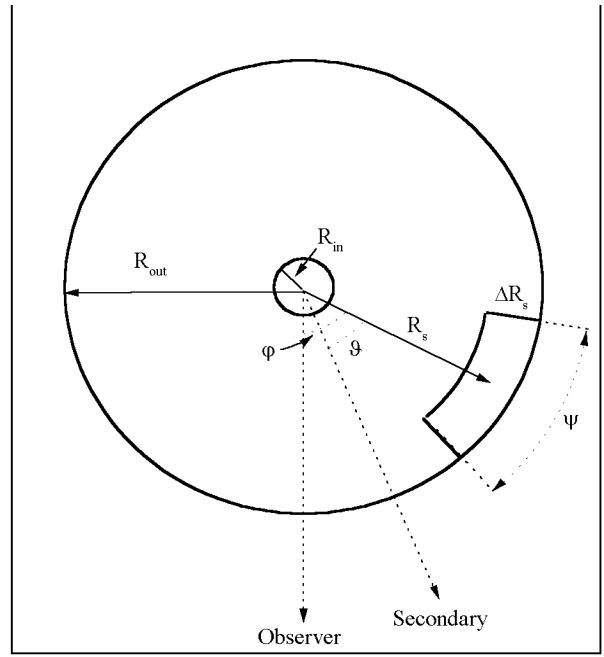,width=8cm,%
bbllx=-2pt,bblly=0pt,bburx=225pt,bbury=255pt,clip=}}}
\caption{Accretion disk geometry for the line profile model described in the
text.}
\label{fig1}
\end{figure}

In close binary systems it is possible to note five basic emission
regions: an accretion disk, a gas stream, a bright spot, a primary and
a secondary components. However, in low--mass systems the accretion disk and
the bright spot only make the main contribution to the radiation of
emission lines (see, for example, Marsh et al., 1990; Marsh and Horne, 1990).
Therefore in our calculations we applied a two--component model which
included a flat Keplerian geometrically thin accretion disk and a bright spot
whose position is constant with respect to the components of a
binary system
(Fig.~\ref{fig1}). We began the modelling of line profiles with calculation of
a symmetrical
double--peaked profile formed in a uniform axisymmetrical disk,
then we added a distorting component formed in the bright spot.

The flat Balmer line decrement usually observed in  spectra of  cataclysmic
variables shows that the hydrogen emission lines are optically thick. In
this case the local emissivity of the lines becomes strongly anisotropic,
because the photons tend to emerge easily in the directions of high
velocity gradients. Therefore for the calculation of the line profiles we
have used the method of Horne and Marsh (1986), taking into account the
Keplerian velocity gradient across the finite thickness of the disk.
To calculate the  emission line profile we divide the disk surface
into a grid of elements, and assign the velocity vector, line strength and
other parameters for each element. The computation of
the profiles proceeds by summing the local line profiles weighted by the
areas of the surface elements. For details see Horne and Marsh (1986) and
Horne (1995).We have assumed a power--law function for  distribution of the
local line emissivity $f(r)$ over the disk's surface $f(r) \propto
r^{-\alpha},$ where r is the radial distance from the disk's centre and
$\alpha=1\div 2.5$ (Smak, 1981; Horne and Saar, 1991).

Free parameters of our model are:
\begin{enumerate}
\item
Parameter $\alpha$;
\item $R=R_{in}$/R$_{out}$ --- the ratio of the inner and the outer radii of
the disk;
\item $V$ --- radial velocity of the outer rim of the accretion disk.

Unfortunately, the theoretical modelling of the stream--disk interaction is in
its infancy. It is known from photometric and spectral studies
that over the
orbital period the bright spot is eclipsed by the outer edge of the
accretion disk (see, for example, Livio et al., 1986).
However, it
is still unclear if the spot is optically thick. By the optically thick
spot
we mean one  for which  anisotropy of the local line emissivity should be
taken into account.  We consider this problem in details in the next section.
In our model we consider  the spot on the accretion disk to have a Keplerian
velocity and to be described by four geometric parameters (Fig.1):  \item
$A$ --- the azimuthal angle of the spot centre relative to the line of sight,
$A=\varphi+\vartheta$ (Fig. 1); \item $\Psi $ --- the spot azimuthal  extent;
\item $R_S$ --- the radial position of the spot centre in
fractions of the outer radius ($R_{out}$=1); \item $\Delta R_S$ --- the
radial extent.

For simplicity we assume that the brightness ratio of the
spot and disk is constant and the spot brightness does not depend on azimuth
($f_S(A)=const)$, and its dependence on radius is described by the function
\item $f_S(r)=B\cdot f(r)\propto B\cdot r^{-\alpha}$, where the free parameter $B$
is the spot brightness. For the further analysis instead of $B$ it is
preferable to use the relative dimensionless luminosity $L$, which is
determined as

\vspace*{0.5cm}
\noindent$
\begin{array}[t]{l} L=\int\limits_{R_{S}-\Delta
R_{S}/2}^{R_{S}+\Delta R_{S}/2}S\cdot B\cdot f(r)\cdot dr= \\ \\
=\frac{\pi}{180}\frac{\Psi \Delta R_{S}B}{2-\alpha}\left[\left(R_{S}+
\frac{\Delta R_{S}}{2}\right)^{2-\alpha}-\left(R_{S}-\frac{\Delta R_{S}}{2}
\right)^{2-\alpha}\right] \end{array} $

\mathstrut
\noindent where  $S$ -- the spot
area, and $B$ -- the spot brightness.
\end{enumerate}
So, our accretion disk model has 8 parameters. Such multiparametric
reverse problems raise a question on the uniqueness and stability of the
solution.
In order to answer it, we analyze how the various parameters of the spot and
disk affect a line profile.
\begin{figure}
{\vbox{\psfig{figure=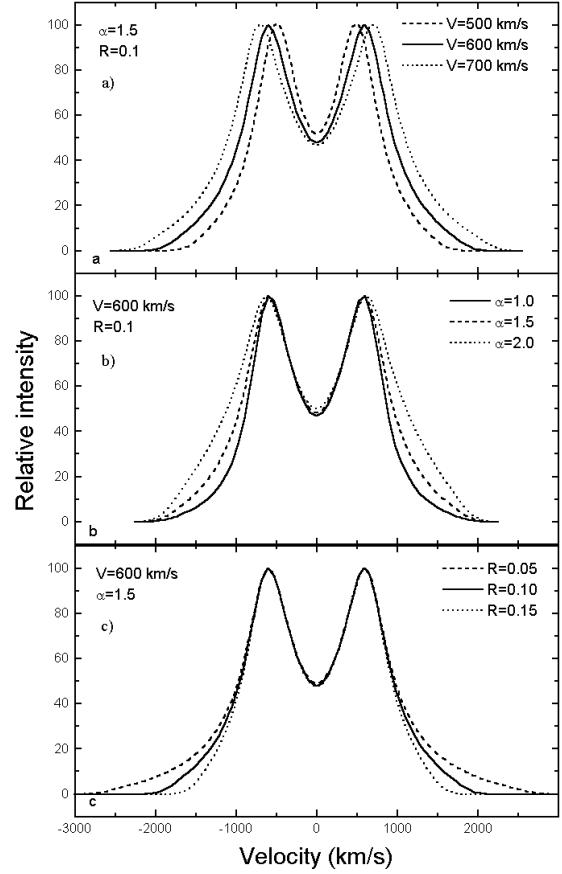,width=8cm,%
bbllx=20pt,bblly=50pt,bburx=200pt,bbury=315pt,clip=}}}
\caption{Relationship between the line profile and the accretion
disk model parameters:  a) V, b) parameter $\alpha$, c) R.}
\label{fig2}
\end{figure}

\section{Line profile dependence on the model parameters}
\subsection{Parameters of the accretion disk}
The dependence of the line profile on the parameters of the uniform accretion
disk is
considered in details by Smak (1981) and Horne and Marsh (1986). They have
shown that the accretion disk parameters basically affect different parts of
line profiles on the
whole, and therefore they can be determined unambiguously. Really,
the velocity of
the outer rim of the accretion disk $V$ defines the  distance between the
peaks in the lines (Fig. 2a), the shape of the line profile depends on the
parameter $\alpha$ (Fig. 2b), and the extent of the wings is determined by
$R$ (Fig. 2c).

\subsection{Parameters of the bright spot}
\begin{figure}[b]
{\vbox{\psfig{figure=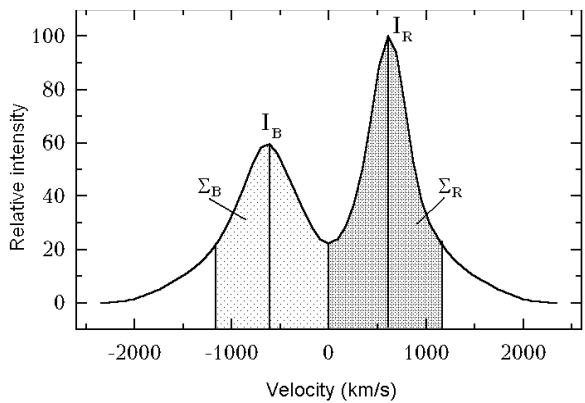,width=8cm,%
bbllx=5pt,bblly=145pt,bburx=180pt,bbury=260pt,clip=}}}
\caption{The areas of the ``humps'', which are used to increase the
accuracy of measuring the asymmetry degree instead of intensities
of line peaks (see text for details).}
\label{fig3}
\end{figure}

When we studied the dependence of the emission line profiles on the
parameters of the bright spot it was important to find out whether it was
possible to make their unambiguous estimates.
The results of testing presented below
are based on the modelling of a series of line profiles of the
accretion disk with the bright spots on different azimuths and having
different parameters. Some of them were fixed here, but others were
changed so that the relative spot luminosity was constant. The calculations
have shown that correct determination of the spot
parameters
depends on its azimuth, which is necessary to find in an independent way.
This is possible from the analysis of the phase variations of asymmetry
degree of emission lines (for example, v/r ratio). The azimuth of the spot
is $%
A=\vartheta +\varphi =1+\varphi -\varphi _{0}$, where $\vartheta $ --- the
phase angle of the spot, $\varphi $ --- the orbital phase, and $\varphi
_{0}$ --- the phase
of the moment when the v/r-ratio is equal to 1. It corresponds to the moment
when
the radial velocity of the S--wave components is zero. Note that in practice
to
improve the  accuracy of measuring the asymmetry degree, instead of the ratio
of intensities of the line peaks it is  preferable to use the quantity:
\begin{description}
\item  $S=\left\{
\begin{array}{l}
\Sigma _{B}/\Sigma _{R} \\
2-\Sigma _{R}/\Sigma _{B}
\end{array}
\right. \hspace{1cm}
\begin{array}{r}
\Sigma _{B}>\Sigma _{R}, \\
\Sigma _{B}<\Sigma _{R}
\end{array}
$
\end{description}
\noindent where S --- the degree of asymmetry, $\Sigma _{B}$ and $\Sigma
_{R}$ ---
the integrals of the line intensity of the violet and the red ``humps'' (or
their
parts in equal ranges of wavelengths), respectively (Fig.~\ref{fig3}).
Moreover, such a quantity is more sensitive to changing the spot parameters.

\subsubsection{Is the spot optically thin or optically thick?}

\begin{figure}[t]
{\vbox{\psfig{figure=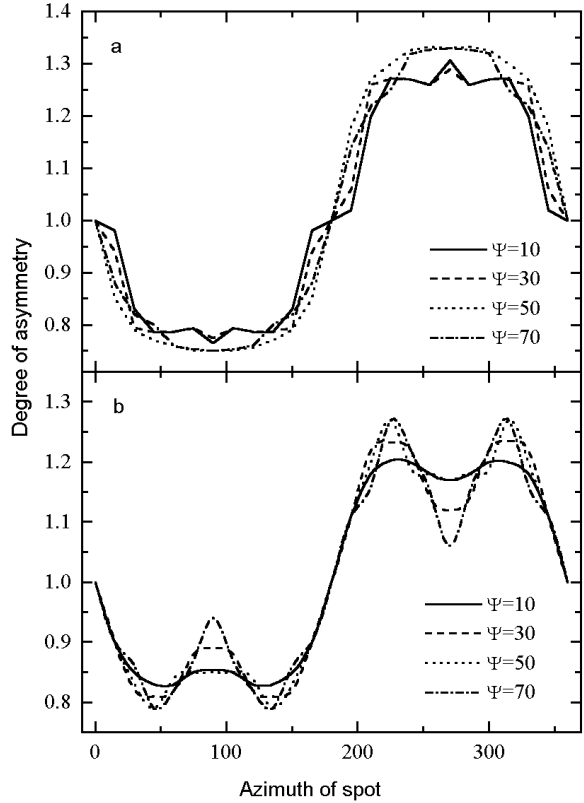,width=8cm,%
bbllx=5pt,bblly=25pt,bburx=210pt,bbury=300pt,clip=}}}
\caption{Asymmetry degree of the emission line, formed in the accretion
disk with the optically thin (a) and optically thick (b) spot
as a function of orbital phase.}
\label{fig4}
\end{figure}

In an optically thick disk the local line emissivity is strongly
anisotropic.  Thus the line surface brightness of the accretion disk must be
enhanced with non--axisymmetrical pattern and proportional to $\left|
sin(2\varphi )\right| $ (Horne and March, 1986; Horne, 1995). This happens
because the velocity gradient on spot azimuths $\pm $45$^{\circ }$ and
$\pm $135$^{\circ}$ is the greatest and the probability of the line photon
tending to emerge is also the highest. So, the observed brightness of
an optically thick spot will vary with its azimuth (due to its limited size).
It will be maximum on azimuths $\pm 45^{\circ }$ and $\pm 135^{\circ }$,
fall on the azimuth $\pm 90^{\circ}$ and minimum on azimuths $0^{\circ }$
and $180^{\circ }$. We have calculated a set of models with different
parameters of the spot at phases covering the full orbital period, and
based on the obtained profiles we have plotted a grid of S--waves
(Fig.~\ref{fig4}). For the optically
thick spot
the S--wave curves are seen to have a depression at spot phase $\pm
$90$^{\circ }$. The  depth of the depression increases with decreasing
azimuth extent of the spot.  Such a depression on the S--wave curve may
suggest that the spot is optically thick, since in the
case of the optically thin spot there is no depression.

\begin{figure}[htbp]
{\vbox{\psfig{figure=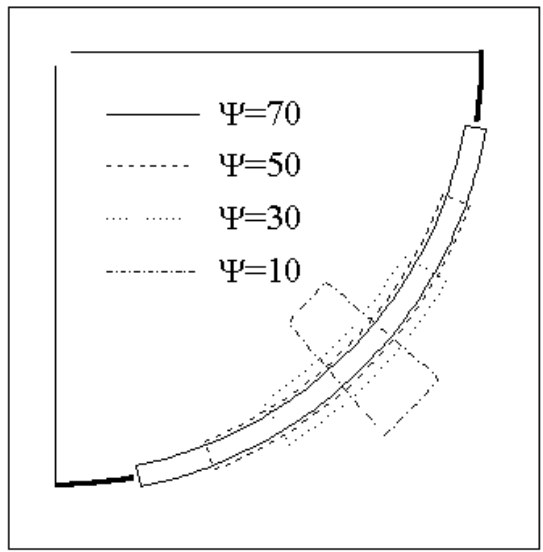,width=8cm,%
bbllx=15pt,bblly=60pt,bburx=105pt,bbury=145pt,clip=}}}
\caption{Spot models used to determine the sensitivity of
the line profiles to the spot shape. All the spots have equal relative
luminosity and equal area.}
\label{fig5}
\end{figure}
\subsubsection{The shape of the spot}

\begin{figure*} {\vbox{\psfig{figure=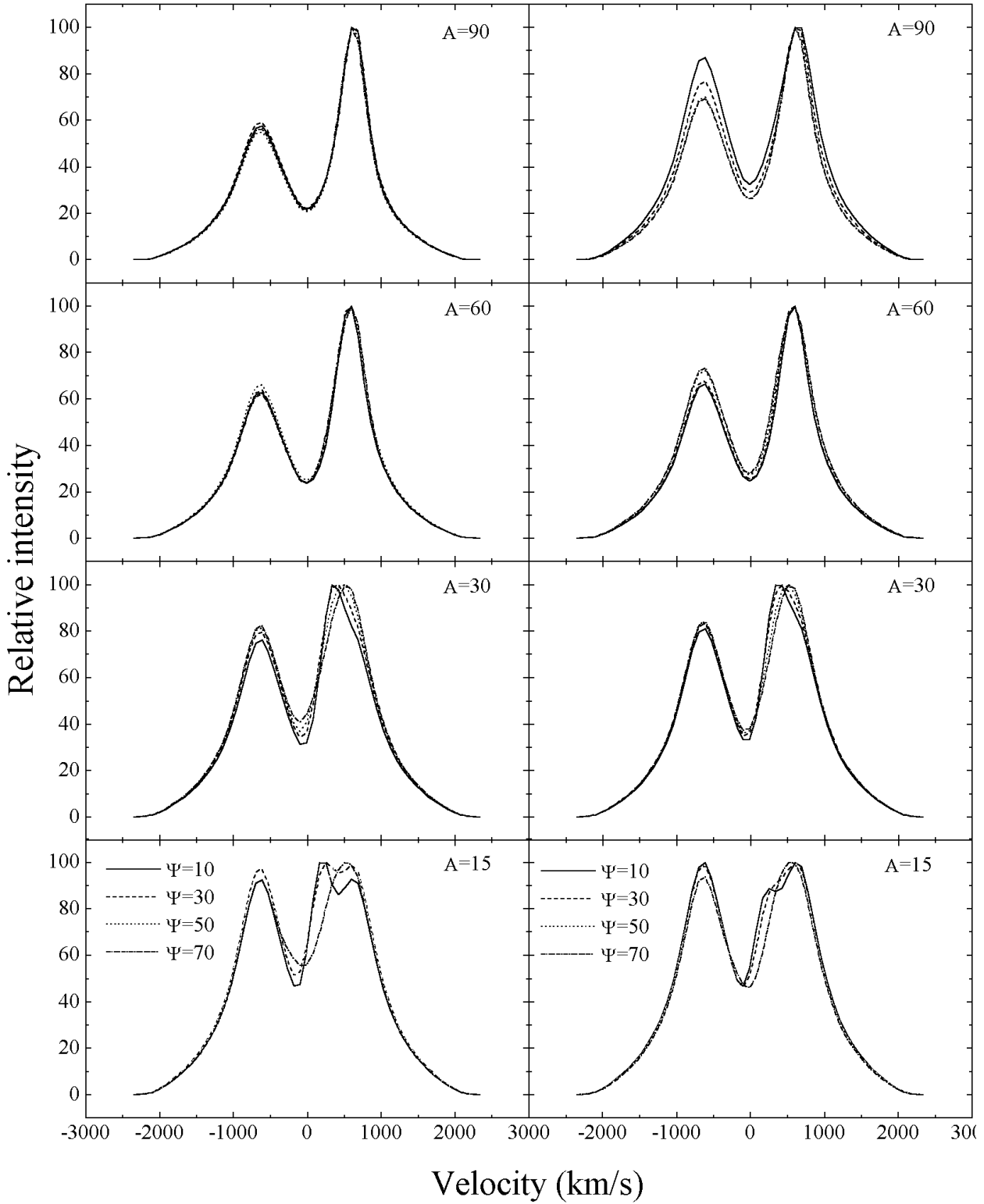,width=16.5cm,%
bbllx=40pt,bblly=140pt,bburx=435pt,bbury=630pt,clip=}}}
\caption{Line profiles as a function of the spot shape. Left ---
optically thin spot, right --- optically thick spot.}
\label{fig6}
\end{figure*}

To find out if the shape of the spot affects the shapes of emission
lines,
we used the models shown in Fig.\,\ref{fig5}. In spots of equal
relative
luminosity and equal area the azimuth extent varied from 10  to 70
degrees. It is seen that the line profile does not depend on the
spot shape for  azimuth $>45\degr \div 60\degr$ and
strongly depends on it at phase $<30\degr\div 45\degr$
(Fig.~\ref{fig6}). This is true both for optically thick and optically thin
spots.

\subsubsection{Radial extent of the spot}

Actually the  radial extent of the spot does not affect the
shape of the line profile. Variation of this parameter over a wide range,
from 0.02 to 0.30, affects slightly the profile at all
phases and any azimuth extent of the spot. This is explained by the fact that
the interval of the radial velocities inside the spot only slightly depends
on $\Delta R_S$.  Therefore it is possible to compensate for the change in
this parameter by changing the spot contrast $B$, i.e. the invariant is the
product $B\cdot \Delta R_S$.  So, to calculate the line profile, it is
necessary to specify the value of $\Delta R_S$ in some way (for example,
the typical radial extent of the bright spot). It is known from photometric
observations of cataclysmic variables that for different systems it lies in
the range  $0.02\div 0.15$ (Rozyczka, 1988).

\subsubsection{Radial position of the  spot on the accretion disk}

\begin{figure*}
{\vbox{\psfig{figure=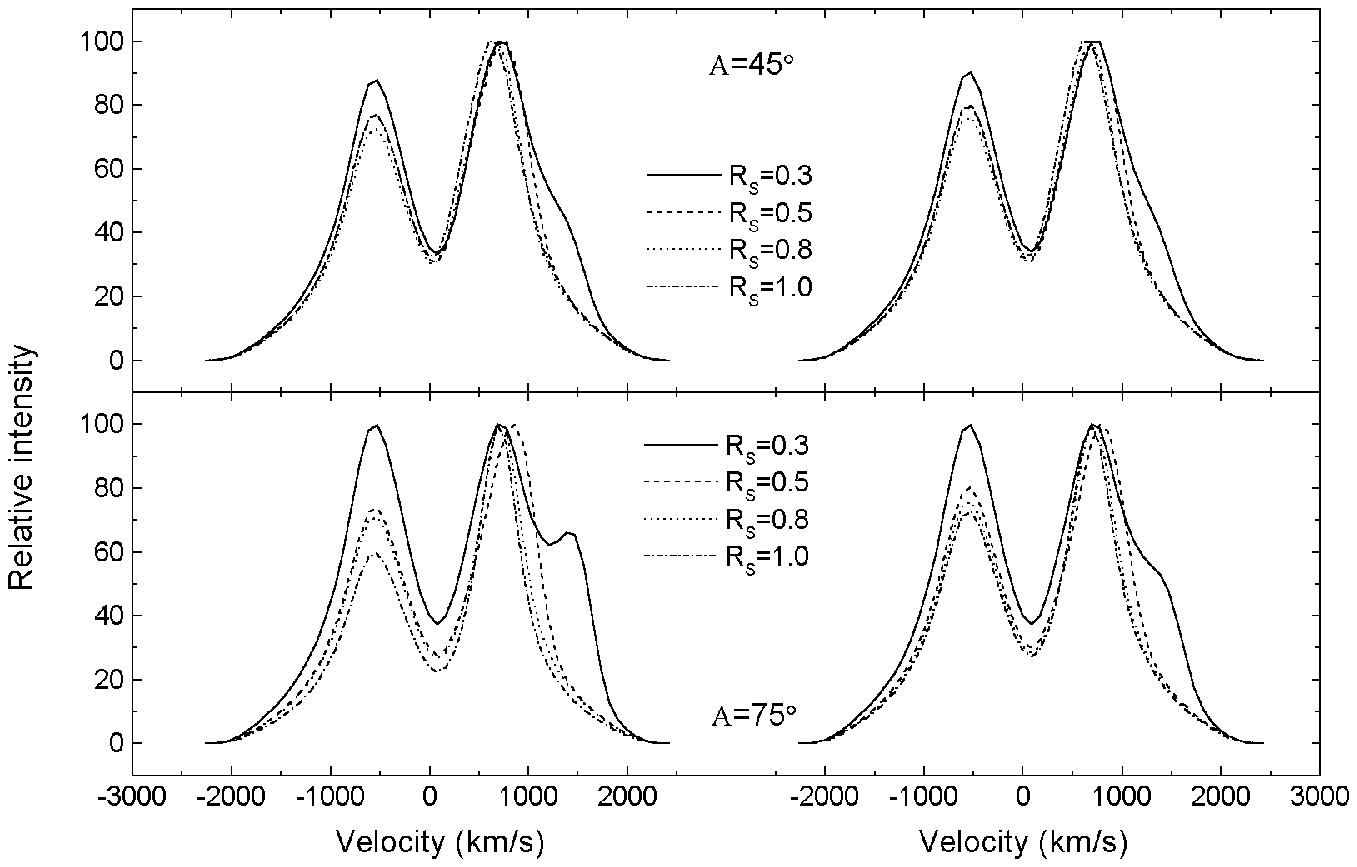,width=16cm,%
bbllx=25pt,bblly=35pt,bburx=415pt,bbury=290pt,clip=}}}
\caption{Relationship between the line profile and  the radial
position of the spot.  The parameters of  optically thin (left) and of
optically thick (right) spots:  $\Psi$=70$^{\circ }$, $\Delta $R$_S$=0.05.}
\label{fig7}
\end{figure*}

\begin{figure*}
{\vbox{\psfig{figure=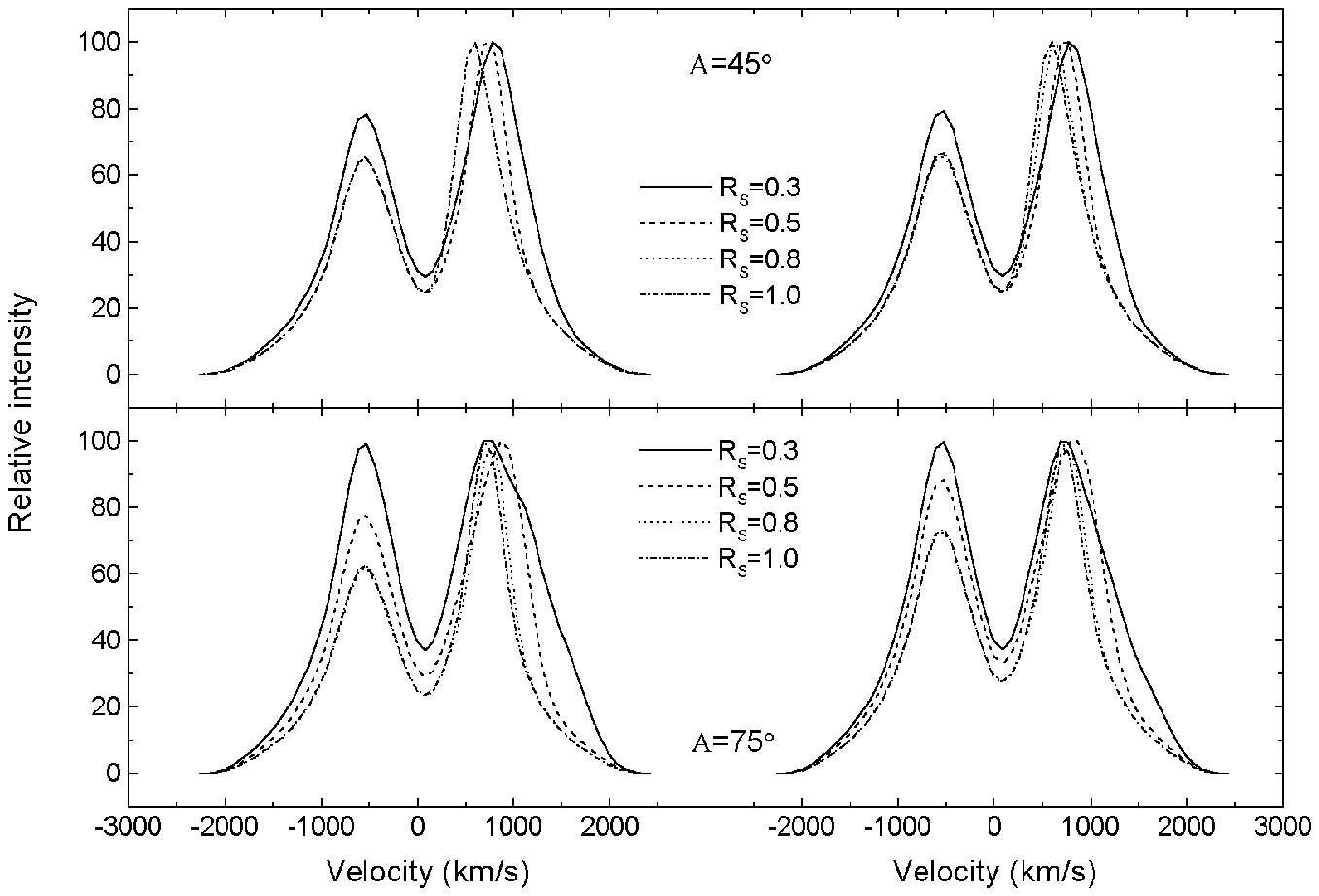,width=16cm,%
bbllx=25pt,bblly=20pt,bburx=415pt,bbury=290pt,clip=}}}
\caption{As in Fig.~\protect\ref{fig7} but the parameters are
$\Psi=10^{\circ }$ and $\Delta R_S=0.34$.}
\label{fig8}
\end{figure*}

This parameter is determined with high confidence for any optically thick
and optically thin spot, which have an azimuth over $45\degr$
(Fig.~\ref{fig7},~\ref{fig8}).


\subsubsection{Radial and azimuthal distributions of the spot brightness}

\begin{figure*}
{\vbox{\psfig{figure=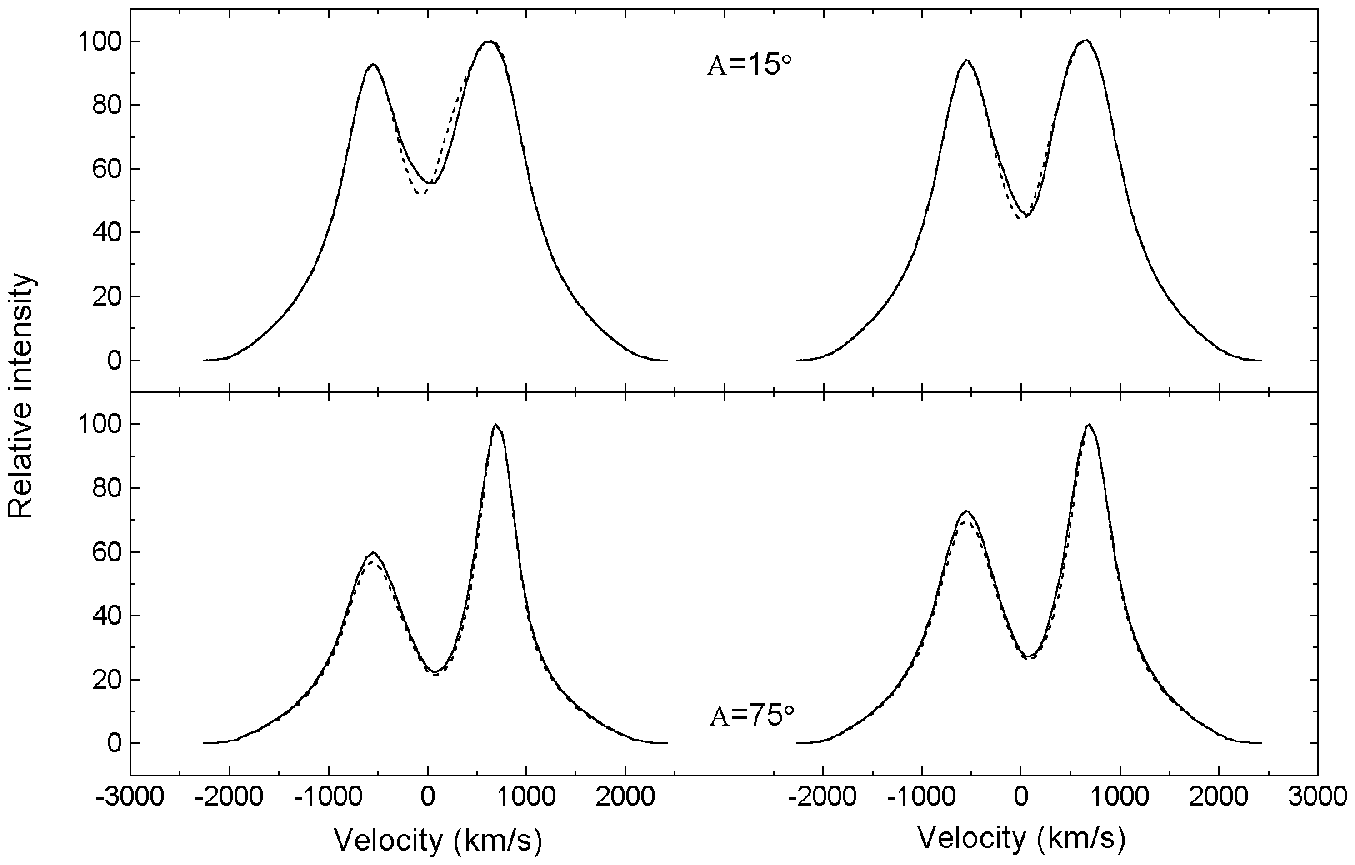,width=16cm,%
bbllx=25pt,bblly=35pt,bburx=415pt,bbury=290pt,clip=}}}
\caption{The line profile vs the azimuthal distribution of the
spot brightness (solid line  --- the spot with uniform distribution, dashed
line --- the spot with asymmetric distribution). The azimuth extent of
optically thin (left) and of  optically thick (right) spots is
$\Psi=70^{\circ }$.}
\label{fig9}
\end{figure*}

\begin{figure*}
{\vbox{\psfig{figure=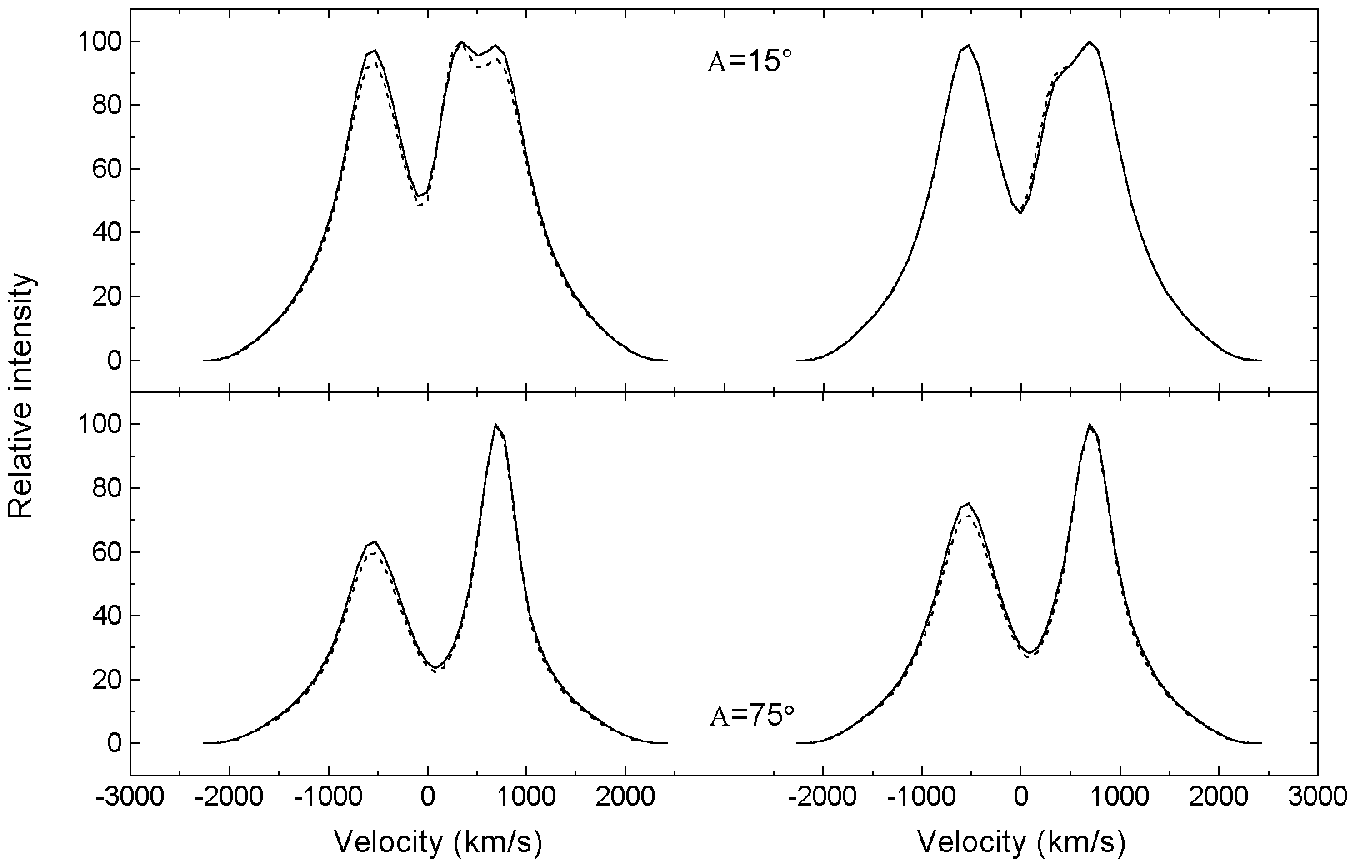,width=16cm,%
bbllx=25pt,bblly=35pt,bburx=415pt,bbury=290pt,clip=}}}
\caption{The same as in Fig.~\protect\ref{fig9} but $\Psi=30\degr$.}
\label{fig10}
\end{figure*}

Calculations have shown that the type of the spot brightness
distribution does not practically influence the line profile. As
an example we adduce the response of the line profile to various
types of the azimuth dependence. We have calculated the models with
asymmetric distribution of the spot brightness (it is decreasing linearly
from $B$ to zero) and with the uniform distribution for comparison.
The relative luminosity of the spot was constant. It is seen
that the minor modifications of the profile appeared for the spots extended
very much in azimuth ($\Psi > 70\degr$), at phases close to
zero (Fig.~\ref{fig9},\ref{fig10}).


\subsection{Gas stream}

The calculation of the line component which is formed in the gas stream has
no principal differences from the modelling of the component formed in the
bright spot. Because the translational velocity of the stream much
exceeds the velocity of its expansion and is highly supersonic, the width of
the local line profile formed in the stream must be much smaller than the
full width of the line formed in  the accretion disk (Lubow and Shu, 1975).
Much of the kinetic energy of the stream is released by radiation
probably at the moment of its collision with the disk. It is known
from observations that usual radii of accretion disks in cataclysmic
variables are $(0.4\div 0.8) R_{L}$, where $R_{L}$ is the Roche lobe
dimension.  The velocity of the stream at such distances from the accreting
star approximately corresponds to Keplerian velocities of the accretion disk
on these radii. For this reason it is complicated to determine the area of
origin of the S--wave component of the observed emission line profile.
However this is still possible to do from analysis of the S--wave, for
example.

\subsection{The secondary component}

Research into some dwarf nova (IP Peg, U Gem) using the Doppler tomography
technique has shown that secondary components in such systems also
contribute to  emission in lines (March and Horne, 1990;
March et al., 1990). However the contribution of their emission to the total
flux
is very small and may be ignored in
calculations.

\section{Evaluation of the accuracy of determination of the parameters}

\begin{figure}
{\vbox{\psfig{figure=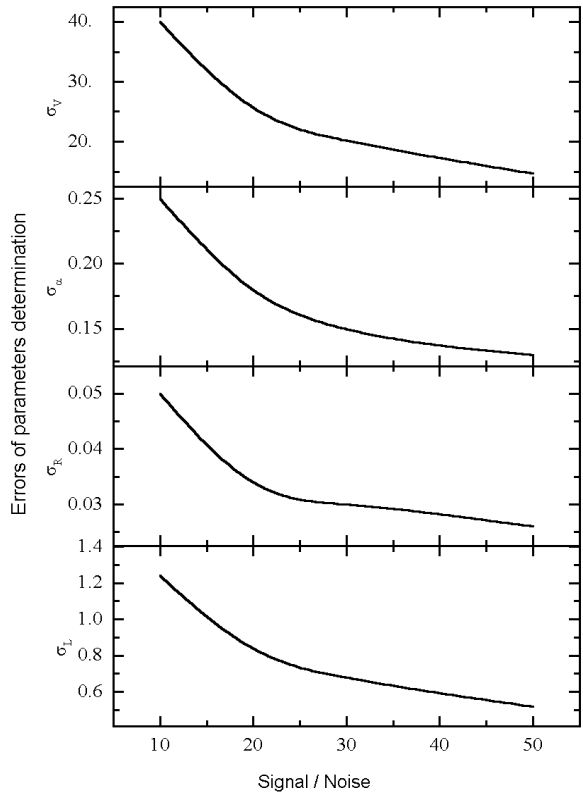,width=8cm,%
bbllx=5pt,bblly=25pt,bburx=210pt,bbury=305pt,clip=}}}
\caption{Errors in determination of parameters vs S/N
ratio.}
\label{fig11}
\end{figure}

For the further analysis of the obtained results it is important to know the
accuracy of determination of the model parameters. It depends on many
factors,
for example, on the method of determination, on the
spectral
resolution, and also on the values of the parameters. Because it is
nearly
impossible to take into account all these factors analytically, we have
decided to use the following statistical method. A line profile calculated
with the known values of the model parameters was normalized so that the
relative intensity of the line was equal to 2 (a value typical of
cataclysmic
variables). Then the profile was distorted by the Poisson noise (the level
of the continuum was varied to  change the S/N ratio). After this
parameters were fitted to the  minimum of  residual deviation of a ``new''
model profile from the ``old'' noisy one. This procedure was repeated
several
hundred times, then the average values and the errors of determination of
the parameters
were estimated. The results of these calculations are shown in
Fig.~\ref{fig11}. As it
can be seen from the plots presented, $V$ and the relative luminosity of the
spot are estimated with the highest accuracy.
For instance, the accuracy of determination of $V$
under
$\rm S/N=30\div 50$ is about 20 km/s. It is better than 5 percent for the
typical
value of $V$ of about 700 km/s. The parameter $\alpha$ is determined quite
confidently. The accuracy of $r$  determination is essentially lower.

\section{Conclusions}
We have presented a technique for calculation of profiles of emission lines
formed in a non--uniform accretion disk. The results of calculations have
shown
that the analysis of spectra obtained at different phases of the orbital
period allows basic parameters of the  spot (such as geometric sizes and
luminosity) to be estimated
and the structure of the accretion disk to be investigated. We
have determined that change in the shape of the emission
line profiles with the variation of different parameters of the  spot
strongly depends on
the azimuth of the spot. Therefore, the necessary condition for  the
accurate
determination of the parameters of the spot is a knowledge of its phase
angle.
By
separating all spectra according to phases of ``the greatest influence'' of
appropriate parameters we can sequentially determine them.

We have found that:
\begin{enumerate}
\item
Analysis of the S--wave allows us to determine the phase angle of the spot
and its
optical depth.
\item The azimuth extent of the spot is determined better on azimuths
$< 30\degr \div 45\degr$, while its radial position is determined better on
azimuths $> 50\degr$.
\item The shape of the line profile is practically insensitive to
modification
in radial extent of the spot. Therefore, for modelling the start value of
this parameter is set by default (it is possible to take, for example,
a typical radial extent of a bright spot). Thus the number of free parameters
of the model decreases by unity.
\end{enumerate}
\begin{acknowledgements}

We thank G.M. Beskin and L.A.\,Pustil'nik for valuable advice and
discussion
of the work. The work was partially supported by the Russian Foundation
of
Basic Research (grant 95-02-03691) and Federal Programme ``Astronomy''.
\end{acknowledgements}

\end{document}